\documentclass[final,1p,times,authoryear]{elsarticle}
\makeatletter
\def\ps@pprintTitle{%
  \let\@oddhead\@empty
  \let\@evenhead\@empty
  \def\@oddfoot{\hfil\thepage\hfil}
  \let\@evenfoot\@oddfoot
}
\makeatother

\usepackage{graphicx}
\usepackage{subcaption}
\usepackage{algorithmicx}%
\usepackage{booktabs}%
\usepackage{algorithm}
\usepackage{algpseudocode}
\usepackage{multirow}
\usepackage{multicol}
\usepackage{float}      
\usepackage{url}
\usepackage{amssymb}
\usepackage{comment}
\usepackage{amsmath}
\usepackage[table]{xcolor}
\usepackage{multirow}

\newcount\Comments  
\usepackage{color}
\definecolor{darkgreen}{rgb}{0,0.5,0}
\definecolor{purple}{rgb}{1,0,1}
\newcommand{\kibitz}[2]{\ifnum\Comments=1\textcolor{#1}{#2}\fi}

\definecolor{orange}{rgb}{1,0.5,0}

\journal{Statistics and Probability Letters}

\begin{document}

\begin{frontmatter}



\title{Infinite hidden Markov models for cylindrical data}

\author[label1,label2]{Federico P. Cortese}
\ead{federico.cortese@unimi.it}

\author[label1,label3]{Luca Rossini}
\ead{luca.rossini@unimi.it}

\affiliation[label1]{
  organization={Department of Economics, Management and Quantitative Methods, 
  \\
  University of Milan},
  country={Italy}
}

\affiliation[label2]{
  organization={
  Institute for Applied Mathematics and Information Technologies, \\
  National Research Council},
  city={Milan},
  country={Italy}
}

\affiliation[label3]{
  organization={Fondazione Eni Enrico Mattei},
  city={Milan},
  country={Italy}
}


\begin{abstract}
We propose an infinite hidden Markov model for cylindrical time series with von Mises-Gamma emissions. Posterior inference is performed using a beam sampler combining conjugate updates and approximate sampling schemes. Simulation studies and two real data applications demonstrate the effectiveness of the proposed methodology.
\end{abstract}




\end{frontmatter}




\section{Introduction}

A growing body of literature studies time series with mixed-type components, which are common in environmental or 
biological applications.
A prominent example is provided by cylindrical data, where each observation consists of a circular and a positive linear component.
Such data naturally arise, for example, when modeling 
wind direction and wind speed or the movement direction and speed of animals
\citep{lagona2015hidden}.
While hidden Markov models (HMMs) have become the standard approach for analyzing cylindrical time series, selecting the appropriate number of latent states remains a fundamental challenge. For example, \citet{holzmann2006hidden} determine this number \emph{a priori}, whereas \citet{zucchini:2017} show that information-criterion-based model selection can lead to conclusions that differ from prior evidence.
From a frequentist perspective, \citet{Harvey2023} 
model wind direction using
score-driven approaches for regime-switching circular time series. 

In this paper, we relax the finite-state assumption by introducing an infinite hidden Markov model \citep[iHMM,][]{beal2001infinite} for cylindrical time series. The iHMM is a Bayesian nonparametric HMM in which the complexity of the latent state space is learned from the data. 
By relying on hierarchical Dirichlet process \citep[HDP,][]{teh2006hierarchical} priors, iHMMs avoid the need to specify the number of hidden states a priori 
while retaining the flexibility to accommodate a broad range of emission distributions.
%
Specifically, the proposal combines a von Mises distribution for the circular component with a Gamma distribution for the positive linear component.
Posterior inference is carried out through an efficient beam sampler \citep{van2008beam},  exploiting conjugate updates whenever available and, following \cite{miller2019fast} and \cite{forbes2015fast}, approximate sampling schemes for the remaining parameters.

Simulation studies demonstrate that the proposed method accurately recovers the true latent structure. The practical usefulness of the approach is illustrated through two real-data applications involving \emph{Drosophila melanogaster} movement and wind speed-direction measurements for the city of Milan.

The 
remainder of the 
paper is organized as follows. Section~\ref{sec:model} introduces the proposed model and posterior sampling scheme. 
Section~\ref{sec:simstud} presents the simulation study, while in Section~\ref{sec:app} 
we apply 
the methodology 
to two real datasets. 
Section~\ref{sec:concl} concludes.
Further details on the sampling scheme, simulation results, and empirical analyses are available in the Supplementary Material.

\section{Model formulation and posterior sampling}
\label{sec:model}

Let $G_0$ be a Dirichlet process \citep[DP,][]{ferguson1973bayesian} with base distribution $H$ and concentration parameter $\gamma$. Using the stick-breaking construction \citep{sethuraman1994constructive}, it admits the representation
$
G_0=\sum_{k=1}^{\infty}\beta_k\delta_{\boldsymbol{\theta}_k},
$
where the weights $\boldsymbol{\beta}=(\beta_1,\beta_2,\ldots)\sim \text{GEM}(\gamma)$ (Griffiths, Engen and McCloskey) satisfy
\[
\beta_k=v_k\prod_{\ell=1}^{k-1}(1-v_\ell),
\qquad
v_k\sim\mathrm{Beta}(1,\gamma),
\qquad k=1,2,\ldots,
\]
and the atoms $\boldsymbol{\theta}_k$ are independently drawn from $H$.
A hierarchical Dirichlet Process (HDP) is obtained by introducing a second-level DP,
$$
\boldsymbol{\pi}_k=(\pi_{k1},\pi_{k2},\ldots)
\sim
\mathrm{DP}(\alpha,\boldsymbol{\beta}),
$$
which defines the $k$-th row of an infinite-dimensional transition matrix. 
The pair $(\boldsymbol{\beta},\{\boldsymbol{\pi}_k\}_k)$ 
forms the main building block of the iHMM, defined as
\begin{equation}
\begin{aligned}
\boldsymbol{\beta}  &\sim \mathrm{GEM}(\gamma), \qquad
\boldsymbol{\pi}_k \sim \mathrm{DP}(\alpha,\boldsymbol{\beta}),\\
s_t \mid s_{t-1}
&\sim \mathrm{Multinomial}(\boldsymbol{\pi}_{s_{t-1}}),\\
\boldsymbol{\theta}_k
&\sim H, \qquad
\boldsymbol{z}_t \mid s_t \sim f(\cdot\mid\boldsymbol{\theta}_{s_t}), \, t=1,\ldots,T,
\end{aligned}
\label{eq:iHMM}
\end{equation}
where $s_t$ denotes the latent state at time $t$, $\boldsymbol{\theta}_k$ contains the emission parameters associated with state $k$, and $f(\cdot\mid\boldsymbol{\theta}_{s_t})$ is the corresponding emission distribution. 
Conditional on the latent state sequence, the observations $\boldsymbol{z}_t$ are therefore  independent.

Let $\boldsymbol{z}_1,\ldots,\boldsymbol{z}_T$ denote a sequence of cylindrical observations, with
$\boldsymbol{z}_t=(x_t,y_t)$, where $x_t\in[0,2\pi)$ is the circular component and $y_t>0$ the positive linear component. Conditional on the latent state, the circular and linear components are assumed independent,
\[
f(\boldsymbol{z}_t\mid s_t=k)
=
f_x(x_t\mid\mu_k,\kappa_k)
f_y(y_t\mid a_k,b_k).
\]
The circular component $f_x$ follows a von Mises distribution with density 
\[
f_x(x_t \mid \mu_k,\kappa_k) \;=\;
\frac{1}{2\pi I_0(\kappa_k)} \exp\!\left\{ \kappa_k \cos(x_t - \mu_k) \right\},
\]
where $\mu_k$ and $\kappa_k$ are the mean direction and concentration, respectively, and $I_0(\cdot)$ denotes the modified Bessel function of the first kind. The positive linear component $f_y$ follows a Gamma density with shape $a_k$ and rate $b_k$.

We assign the following independent priors to the emission parameters of the linear and circular components: 
$
a_k\sim\mathrm{Gamma}(\delta,\eta),
$ $
b_k\sim\mathrm{Gamma}(\zeta,\xi),
$
and
$
\mu_k\mid\kappa_k
\sim
\mathrm{VM}(\mu_0,\kappa_0),$
$
\pi(\kappa_k)
\propto
I_0(\kappa_k)^{-\rho}
e^{-\chi\kappa_k}.
$
We use weakly informative priors by setting $\mu_0=0$ and all remaining emission hyperparameters equal to 1.
The prior distributions for the HDP concentration parameters are $\alpha\sim \text{Gamma}(a_\alpha,b_\alpha)$ and $\gamma\sim \text{Gamma}(a_\gamma,b_\gamma)$, where $a_\alpha=a_\gamma=1$, and $b_\alpha=b_\gamma=5$.

Posterior inference is performed using the beam sampler of \citet{van2008beam}, a method that combines slice sampling and the forward-filtering backward sampling algorithm \citep[FFBS,][]{chib1996calculating}. At each iteration, an auxiliary slice variable $u_t\sim\mathcal{U}(0,\pi_{s_{t-1},s_{t}})$ is introduced to dynamically truncate the infinite state space to a finite set of feasible states. Conditional on these variables, the latent state sequence is sampled exactly using the FFBS algorithm, while the original iHMM is recovered after marginalizing out the slice variables.

Given the updated state sequence, the transition probability vectors $\boldsymbol{\pi}_k$ and the global 
weights $\boldsymbol{\beta}$ are sampled using the standard Gibbs updates for HDP, followed by updates of the concentration parameters $\alpha$ and 
$\gamma$ \citep{teh2006hierarchical}.

Finally, the emission parameters $\boldsymbol{\theta}_k=(a_k,b_k,\mu_k,\kappa_k)$, $k=1,\ldots,K$, are updated independently across states. For the Gamma distribution, the rate parameter $b_k$ admits a conjugate Gamma full conditional, 
\begin{equation}
\label{eq:full_bk}
b_k|a_k,y_1,\ldots,y_T,\boldsymbol{s}
\sim
\text{Gamma}\left(
\zeta+n_k a_k,
\xi+\sum_{t:s_t=k}y_t
\right),
\end{equation}
where $n_k$ is the number of observations allocated in state $k$. The full conditional distribution of the shape parameter $a_k$ is not available in closed form,  so it is sampled using the efficient Gamma approximation of \citet{miller2019fast}. For the circular component, the conditional posterior of $\mu_k$ remains a von Mises distribution, 
\begin{equation}
\label{eq:full_muk}
\mu_k \mid \kappa_k, x_1,\ldots,x_T,\boldsymbol{s}
\sim
\mathrm{VM}\left(\operatorname{atan2}(S_k,C_k), \sqrt{C_k^2+S_k^2}\right),
\end{equation}
where
$
C_k=\kappa_k\sum_{t:s_t=k}\cos x_t+\kappa_0\cos\mu_0,
$
$
S_k=\kappa_k\sum_{t:s_t=k}\sin x_t+\kappa_0\sin\mu_0$. 
The concentration parameter $\kappa_k$ is sampled using the shifted-Gamma rejection sampler of \citet{forbes2015fast}.
Algorithm~\ref{alg:beam_sampler} summarizes the proposed sampling scheme.

\begin{algorithm}[H]
\caption{Beam sampler for cylindrical iHMM}
\label{alg:beam_sampler}
\begin{algorithmic}[1]
\State Initialize $\boldsymbol{s}$, $\boldsymbol{\pi}$, $\boldsymbol{\beta}$, $\{a_k,b_k,\mu_k,\kappa_k\}_k$, $\alpha$, $\gamma$
\For{$i=1,\ldots,N$}
    \State Sample $u_t\sim\mathcal U(0,\pi_{s_{t-1},s_t})$, $t=1,\ldots,T$
    \State Sample $\boldsymbol{s}$ by FFBS subject to $\pi_{s_{t-1},s_t}>u_t$
    \State Sample $\boldsymbol{\beta}\mid\boldsymbol{s},\gamma$
    and $\boldsymbol{\pi}\mid\boldsymbol{s},\boldsymbol{\beta},\alpha$ 
    \State Sample $\{a_k,b_k\}_k$ using
    \eqref{eq:full_bk} and \cite{miller2019fast}
    \State Sample $\{\mu_k,\kappa_k\}_k$ using 
    \eqref{eq:full_muk} and \cite{forbes2015fast}
    \State Sample $\alpha,\gamma$ 
\EndFor
\end{algorithmic}
\end{algorithm}

\section{Simulation study}
\label{sec:simstud}

To evaluate the proposed methodology, we consider different simulation settings obtained by varying time series length $T\in\{100,500,1\,000\}$ and true number of latent states $K\in\{2,4\}$. 
The latent state sequence is generated from a Markov chain with self-transition probability $\pi_{ii}\in\{0.70,0.95\}$, and off-diagonal probabilities $\pi_{ij}=(1-\pi_{ii})/(K-1)$, $i,j=1,\ldots,K$, $i\neq j$. Conditional on the latent state, the circular component is generated from a von Mises distribution and the linear component from a Gamma distribution. 
The emission parameters are chosen to resemble those observed in the empirical applications of Section~\ref{sec:app}. 
Specifically,

\[
\begin{array}{c|cccc}
K & ({\mu}_1,\ldots,\mu_k) & ({\kappa}_1,\ldots,\kappa_k) & ({a}_1,\ldots,a_k) & ({b}_1,\ldots,b_k) \\
\hline
2 &
(0,\pi/2) &
(10,0.5) &
(4,4) &
(2,8) \\[0.3em]
4 &
(3\pi/2,\pi,\pi/2,0) &
(1,2,0.5,10) &
(4,4,4,4) &
(8,2,4,1.33)
\end{array}
\]
For each combination of $\{K, T, \pi_{ii}\}$, we generate 200 datasets and fit the proposed model, retaining $50\,000$ posterior samples after $1\,000$ burn-in iterations.
We assess performance in terms of adjusted Rand index (ARI), computed between the true and estimated latent state sequences. We also report the \cite{Geweke1992} success rate, defined as the proportion of parameters for which the corresponding test-statistic lies in $[-2,2]$, and the median autocorrelation time (ACT).

The results in Table~\ref{tab:simres} show high clustering accuracy and satisfactory convergence across most scenarios. Performance deteriorates only in the most challenging scenario ($T=100$, $K=4$), especially under low state persistence. This behavior is expected because accurate estimation of all state-conditional parameters requires each latent state to be visited sufficiently often, which is less likely when the time series is short and the number of states is relatively large.
ACT values are generally higher for lower $\pi_{ii}$ and larger $K$. Overall, the results suggest that a thinning interval of approximately $10$ iterations is sufficient to substantially reduce autocorrelation.

\begin{table}[t]
\centering
\caption{Median across 200 replications of the adjusted Rand index (ARI), Geweke success rate, and autocorrelation time (ACT), with interquartile range in parentheses.}
\label{tab:simres}
\resizebox{\textwidth}{!}{
\begin{tabular}{rcccccccccccc}
\toprule
&
\multicolumn{6}{c}{$\pi_{ii}=0.95$} &
\multicolumn{6}{c}{$\pi_{ii}=0.70$} \\
\cmidrule(lr){2-7}\cmidrule(lr){8-13}
&
\multicolumn{3}{c}{$K=2$} &
\multicolumn{3}{c}{$K=4$} &
\multicolumn{3}{c}{$K=2$} &
\multicolumn{3}{c}{$K=4$} \\
\cmidrule(lr){2-4}\cmidrule(lr){5-7}
\cmidrule(lr){8-10}\cmidrule(lr){11-13}
$T$
& ARI & Geweke & ACT
& ARI & Geweke & ACT
& ARI & Geweke & ACT
& ARI & Geweke & ACT \\
\midrule

100
&
\shortstack{0.960\\{\scriptsize (0.046)}}
&
\shortstack{1.000\\{\scriptsize (0.100)}}
&
\shortstack{1.427\\{\scriptsize (0.598)}}
&
\shortstack{0.898\\{\scriptsize (0.239)}}
&
\shortstack{0.900\\{\scriptsize (0.200)}}
&
\shortstack{2.701\\{\scriptsize (3.510)}}
&
\shortstack{0.845\\{\scriptsize (0.110)}}
&
\shortstack{1.000\\{\scriptsize (0.100)}}
&
\shortstack{2.223\\{\scriptsize (0.584)}}
&
\shortstack{0.374\\{\scriptsize (0.271)}}
&
\shortstack{0.933\\{\scriptsize (0.200)}}
&
\shortstack{5.134\\{\scriptsize (6.385)}} \\[0.3em]

500
&
\shortstack{0.976\\{\scriptsize (0.016)}}
&
\shortstack{0.900\\{\scriptsize (0.100)}}
&
\shortstack{1.661\\{\scriptsize (0.590)}}
&
\shortstack{0.928\\{\scriptsize (0.059)}}
&
\shortstack{0.850\\{\scriptsize (0.225)}}
&
\shortstack{4.636\\{\scriptsize (2.709)}}
&
\shortstack{0.846\\{\scriptsize (0.039)}}
&
\shortstack{1.000\\{\scriptsize (0.100)}}
&
\shortstack{2.179\\{\scriptsize (0.492)}}
&
\shortstack{0.595\\{\scriptsize (0.093)}}
&
\shortstack{0.900\\{\scriptsize (0.254)}}
&
\shortstack{9.927\\{\scriptsize (7.145)}} \\[0.3em]

1\,000
&
\shortstack{0.972\\{\scriptsize (0.012)}}
&
\shortstack{0.900\\{\scriptsize (0.200)}}
&
\shortstack{1.808\\{\scriptsize (0.635)}}
&
\shortstack{0.931\\{\scriptsize (0.039)}}
&
\shortstack{0.850\\{\scriptsize (0.200)}}
&
\shortstack{5.333\\{\scriptsize (3.957)}}
&
\shortstack{0.850\\{\scriptsize (0.031)}}
&
\shortstack{1.000\\{\scriptsize (0.100)}}
&
\shortstack{2.139\\{\scriptsize (0.315)}}
&
\shortstack{0.643\\{\scriptsize (0.046)}}
&
\shortstack{0.850\\{\scriptsize (0.250)}}
&
\shortstack{12.925\\{\scriptsize (6.207)}} \\

\bottomrule
\end{tabular}
}
\end{table}

\section{Applications}
\label{sec:app}

In this section, we apply the proposed methodology to two real-data applications involving \textit{Drosophila melanogaster} movement and wind measurements. For both datasets, we run ten independent chains, each retaining $50\,000$ posterior samples after $1\,000$ burn-in iterations and applying thinning every $10$ iterations.
To improve convergence, following \citet{cortese2026comparison} we initialize the latent state sequence through partitioning around medoids based on an equally weighted convex combination of the \cite{gower1971general} dissimilarity and the circular distance.
For ease of interpretation, the state-conditional Gamma distributions are summarized in terms of their means, $\lambda_k=a_k/b_k$.

\subsection{Drosophila melanogaster movement}

We fit an iHMM to the \textit{Drosophila melanogaster} movement dataset, a well-known benchmark for hidden Markov models with cylindrical observations \citep{zucchini:2017}. 
The data consist of successive measurements of movement direction and speed for two \textit{Drosophila} genotypes, one wild-type ($w$) and one mutant ($m$), with \(T=180\) and \(T=381\) consecutive observations, respectively. The aim is to characterize subtle changes in locomotion induced by genetically targeted alterations of the underlying neural circuitry \citep{holzmann2006hidden}. Conditionally on the latent state, movement direction and speed are modeled as independent von Mises and Gamma distributions, respectively.
Behavioral studies indicate that \textit{Drosophila} locomotion is primarily characterized by two distinct movement patterns: \emph{forward peristalsis}, corresponding to relatively fast and persistent motion, and \emph{head swinging and turning}, associated with slower movement and greater directional variability \citep{suster2003targeted}. This biological evidence, however, is not fully reflected by finite-state HMMs. While \citet{holzmann2006hidden} assumed two hidden states \emph{a priori}, \citet{zucchini:2017} reported that both AIC and BIC selected a three-state model.

The posterior distribution concentrates on two behavioral states for the wild-type larvae and three states for the mutant larvae. For the wild type, the two regimes are characterized by posterior median mean speeds of $\boldsymbol{\lambda}^{(w)}=(0.51,1.02)$ and concentration parameters $\boldsymbol{\kappa}^{(w)}=(1.52,7.09)$, with mean directions $\boldsymbol{\mu}^{(w)}=(0.29,0.07)$. 
In contrast, the mutant exhibits three distinct regimes with posterior median mean speeds $\boldsymbol{\lambda}^{(m)}=(0.36,0.22,0.08)$, concentration parameters $\boldsymbol{\kappa}^{(m)}=(4.16,0.66,0.34)$, and mean directions $\boldsymbol{\mu}^{(m)}=(6.18,6.06,2.98)$ radians. 
These results indicate that the mutant exhibits an additional intermediate-speed regime, highlighting locomotor differences between the two genotypes.
Posterior sampling exhibits satisfactory convergence, with Geweke success rates of $1.00$ and $0.93$ for the wild-type and mutant datasets, respectively, and median ACTs of $1.61$ and $2.26$.

\subsection{Wind dynamics}
Wind speed and direction constitute a typical example of cylindrical data and are routinely collected in environmental monitoring \citep{lagona2015hidden}. Joint modeling of these variables is relevant in several applications, including wildfire risk assessment, air quality analysis and energy market forecasting.
We analyze daily average wind speed and direction in the city of Milan recorded by ARPA Lombardia\footnote{\url{https://www.arpalombardia.it/dati-e-indicatori/}} from January 1, 2020, to May 20, 2023, for a total of $T=1,236$ observations.

The posterior distribution consistently selects five states. 
Table~\ref{tab:wind} reports the posterior medians and 95\% credible intervals for the state-conditional emission parameters. State~1 is characterized by the highest average wind speed and a relatively concentrated south-easterly direction ($\mu=5.64$ rad). States~2 and~3 correspond to the weakest winds and exhibit highly dispersed directions. 
Finally, States~4 and~5 have similar average wind speeds but markedly different directional concentrations, with the latter highly concentrated around $1.66$ rad, corresponding to northerly winds.
\begin{table}[ht]
\centering
\setlength{\tabcolsep}{8pt}
\renewcommand{\arraystretch}{1.3}
\caption{Posterior medians and 95\% credible intervals for the emission parameters.}
\label{tab:wind}
\begin{tabular}{lccccc}
\midrule
& State 1 & State 2 & State 3 & State 4 & State 5\\
\midrule

$\mu$
&
\begin{tabular}{@{}c@{}}
{\footnotesize 5.640}\\[-1.1ex]
{\scriptsize (5.410, 5.897)}
\end{tabular}
&
\begin{tabular}{@{}c@{}}
{\footnotesize 4.627}\\[-1.1ex]
{\scriptsize (0.056, 6.225)}
\end{tabular}
&
\begin{tabular}{@{}c@{}}
{\footnotesize 3.494}\\[-1.1ex]
{\scriptsize (2.008, 4.126)}
\end{tabular}
&
\begin{tabular}{@{}c@{}}
{\footnotesize 3.276}\\[-1.1ex]
{\scriptsize (1.761, 3.665)}
\end{tabular}
&
\begin{tabular}{@{}c@{}}
{\footnotesize 1.659}\\[-1.1ex]
{\scriptsize (1.599, 1.717)}
\end{tabular}
\\[-1.1ex]

$\kappa$
&
\begin{tabular}{@{}c@{}}
{\footnotesize 2.171}\\[-1.1ex]
{\scriptsize (1.428, 3.234)}
\end{tabular}
&
\begin{tabular}{@{}c@{}}
{\footnotesize 0.209}\\[-1.1ex]
{\scriptsize (0.013, 0.560)}
\end{tabular}
&
\begin{tabular}{@{}c@{}}
{\footnotesize 0.269}\\[-1.1ex]
{\scriptsize (0.099, 0.611)}
\end{tabular}
&
\begin{tabular}{@{}c@{}}
{\footnotesize 0.650}\\[-1.1ex]
{\scriptsize (0.166, 0.831)}
\end{tabular}
&
\begin{tabular}{@{}c@{}}
{\footnotesize 7.087}\\[-1.1ex]
{\scriptsize (5.524, 9.027)}
\end{tabular}
\\[-1.1ex]

$\lambda$
&
\begin{tabular}{@{}c@{}}
{\footnotesize 2.183}\\[-1.1ex]
{\scriptsize (1.864, 2.574)}
\end{tabular}
&
\begin{tabular}{@{}c@{}}
{\footnotesize 0.412}\\[-1.1ex]
{\scriptsize (0.291, 0.741)}
\end{tabular}
&
\begin{tabular}{@{}c@{}}
{\footnotesize 0.727}\\[-1.1ex]
{\scriptsize (0.394, 1.053)}
\end{tabular}
&
\begin{tabular}{@{}c@{}}
{\footnotesize 1.060}\\[-1.1ex]
{\scriptsize (0.422, 1.090)}
\end{tabular}
&
\begin{tabular}{@{}c@{}}
{\footnotesize 1.259}\\[-1.1ex]
{\scriptsize (1.209, 1.316)}
\end{tabular}
\\

\bottomrule
\end{tabular}
\end{table}

\begin{figure}[t]
\centering
\includegraphics[width=\linewidth]{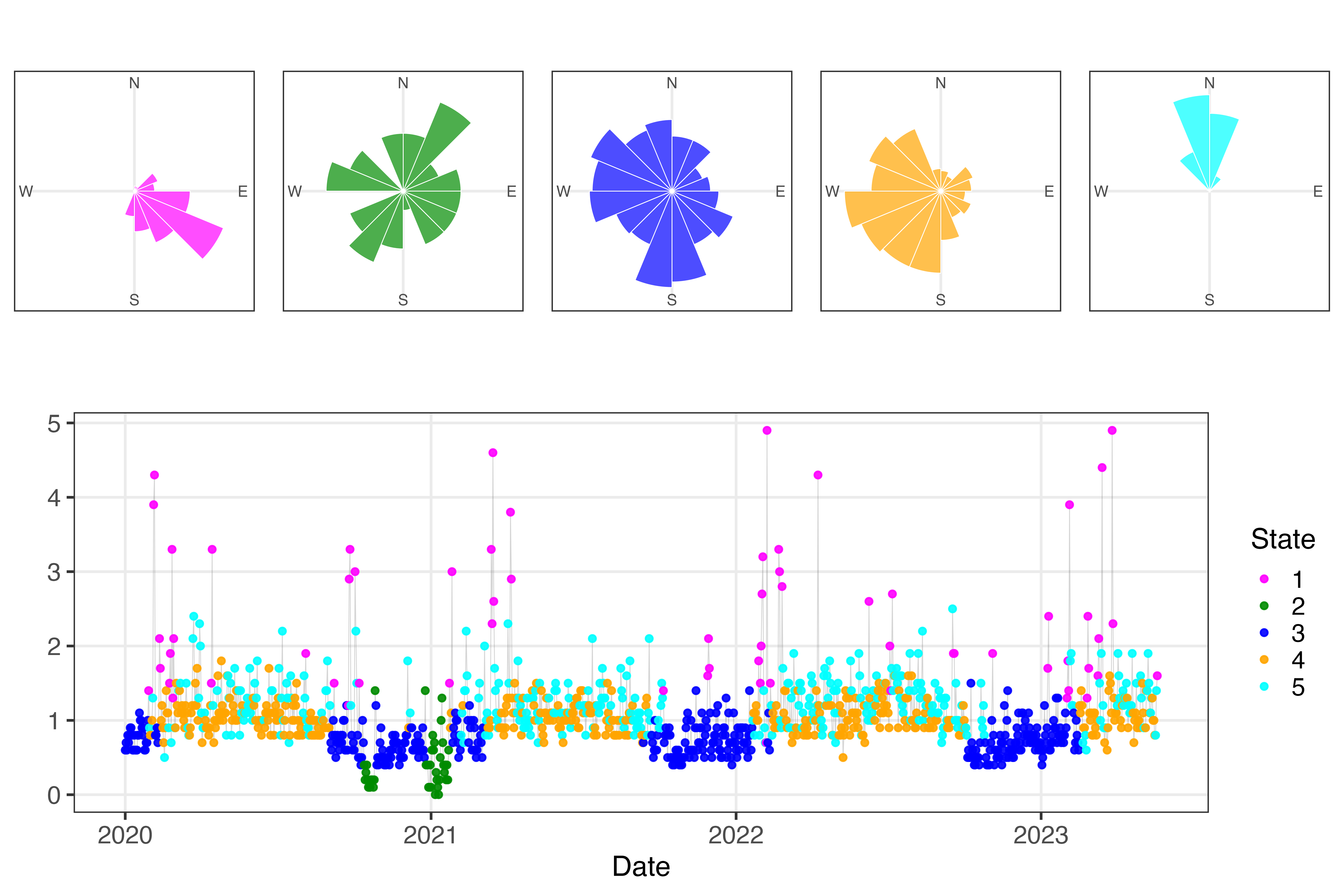}
\caption{Daily wind speed (bottom) and state-conditional distributions of wind direction (top), colored according to the inferred latent states.}
\label{fig:wind}
\end{figure}

Figure~\ref{fig:wind} displays the time series of wind speed with the maximum \emph{a posteriori} state sequence and the corresponding state-conditional wind direction distributions.

Posterior inference exhibits satisfactory convergence and mixing, with a Geweke success rate of $0.92$ and median ACT of $1.54$.

\section{Conclusions}
\label{sec:concl}
We proposed an infinite hidden Markov model for cylindrical time series. The proposed approach overcomes the need to specify or select the number of latent states, while retaining interpretable state-specific emission distributions. Simulation studies and real-data applications demonstrated the effectiveness of the methodology and its practical applicability. 
Future work may consider copula-based dependence between the linear and circular components \citep{lagona2019copula}, as well as state persistence through infinite hidden semi-Markov models to improve posterior inference.


\clearpage

\renewcommand{\thesection}{S\arabic{section}}
\setcounter{section}{0}

{\huge \centering Supplementary Material}

\vspace{3mm}

\noindent
Section~\ref{sec:sampler} provides additional details on the posterior sampling scheme for the emission parameters 
characterizing the circular and linear components. Section~\ref{sec:sim} presents additional results from the simulation study of the main paper. Finally, Section~\ref{sec:emp} reports supplementary tables and figures for the two empirical applications presented in Section~4 of the main paper.

\section{Posterior sampling of the emission distribution parameters}
\label{sec:sampler}

Let $\boldsymbol{z}_1,\ldots,\boldsymbol{z}_T$ be a collection of $T$ cylindrical observations with components
$
\boldsymbol{z}_t=(x_t,y_t),
$
where $x_t\in[0,2\pi)$ is the circular component and $y_t>0$ is the positive linear component.
We define the infinite hidden Markov model as
\begin{equation}
\begin{aligned}
\boldsymbol{\beta} &\sim \mathrm{GEM}(\gamma),\\
\boldsymbol{\pi}_k &\sim \mathrm{DP}(\alpha,\boldsymbol{\beta}),\, k=1,\ldots,K,\\
\boldsymbol{\theta}_k &\sim H,\\
s_t\mid s_{t-1}
&\sim
\mathrm{Multinomial}(\boldsymbol{\pi}_{s_{t-1}}),\\
\boldsymbol{z}_t\mid s_t
&\sim
f(\cdot\mid\boldsymbol{\theta}_{s_t}),
\end{aligned}
\label{eq:iHMM}
\end{equation}
where
\[
\boldsymbol{\theta}_k=(a_k,b_k,\mu_k,\kappa_k).
\]
Conditionally on the latent state, the linear and circular components are assumed independent,
\[
f(\boldsymbol{z}_t\mid s_t=k)
=
f_x(x_t\mid\mu_k,\kappa_k)f_y(y_t\mid a_k,b_k).
\]

\subsection{Linear part}

The positive linear component is modeled as
\begin{align}
y_t\mid s_t=k
\sim
\mathrm{Gamma}(a_k,b_k),
\label{eq:y_t}
\end{align}

where $a_k$ and $b_k$ denote the shape and rate parameters, respectively.

We assign independent Gamma priors

\[
a_k\sim\mathrm{Gamma}(\delta,\eta),
\qquad
b_k\sim\mathrm{Gamma}(\zeta,\xi).
\]

\subsubsection*{Sampling of $a_k$}

The full conditional distribution of $a_k$ is not available in closed form.
Following \citet{miller2019fast}, we temporarily reparameterize the Gamma distribution in terms of its mean,
\[
\lambda_k=\frac{a_k}{b_k},
\]
so that
\[
y_t\mid s_t=k
\sim
\mathrm{Gamma}\!\left(a_k,\frac{a_k}{\lambda_k}\right).
\]
Conditionally on $\lambda_k$, the full conditional of $a_k$ is approximated by a Gamma distribution,
\[
p(a_k\mid\{y_t:s_t=k\},\lambda_k,\delta,\eta)
\approx
\mathrm{Gamma}(A_k,B_k).
\]
The approximation is obtained by matching the first and second derivatives of the log-Gamma density with those of the exact log-full conditional at a point close to its mean. Starting from an initial Gamma approximation, the parameters $A_k$ and $B_k$ are updated iteratively, where at each iteration the matching point is taken as the current Gamma mean,
\[
a_k=\frac{A_k}{B_k}.
\]
The iterations are repeated until convergence, yielding the parameters $(A_k,B_k)$ of the approximating Gamma distribution, from which one may sample
\[
a_k
\sim
\mathrm{Gamma}(A_k,B_k),
\]
as an independent proposal within a Metropolis--Hastings step. Algorithm~\ref{alg:shape-conditional} summarizes the approximation procedure.

\begin{algorithm}[H]
\caption{Approximation of the full conditional of $a_k$ \citep{miller2019fast}}
\label{alg:shape-conditional}
\begin{algorithmic}[1]
\Require
Data $y_1,\ldots,y_T$, parameters $\lambda_k,\delta,\eta>0$, tolerance $\epsilon>0$, maximum number of iterations $M$. $\psi$ and $\psi'$ denote the digamma and trigamma functions.
\State $n_k=\#\{t:s_t=k\}$
\State $R\gets\sum_{t:s_t=k}\log(y_t)$
\State $W\gets\sum_{t:s_t=k}y_t$
\State $V\gets W/\lambda_k-R+n_k\log(\lambda_k)-n_k$
\State $A_k\gets\delta+n_k/2$
\State $B_k\gets\eta+V$
\For{$j=1,\ldots,M$}
\State $a_k\gets A_k/B_k$
\State $A_k\gets\delta-n_ka_k+n_ka_k^2\psi'(a_k)$
\State $B_k\gets\eta+(A_k-\delta)/a_k-n_k\log(a_k)+n_k\psi(a_k)+V$
\If{$\left|a_k/(A_k/B_k)-1\right|<\epsilon$}
\State \Return $A_k,B_k$
\EndIf
\EndFor
\State \Return $A_k,B_k$ 
\end{algorithmic}
\end{algorithm}

\subsubsection*{Sampling of $b_k$}

Since a Gamma prior is assigned to $b_k$, its full conditional is conjugate,
\[
b_k\mid a_k,y_1,\ldots,y_T,\boldsymbol{s}
\sim
\mathrm{Gamma}
\left(
\zeta+n_ka_k,
\xi+\sum_{t:s_t=k}y_t
\right).
\]
Indeed, conditionally on $a_k$ and on the observations allocated to state $k$, the likelihood contribution involving $b_k$ is
\[
\prod_{t:s_t=k}
\frac{b_k^{a_k}}{\Gamma(a_k)}
y_t^{a_k-1}
\exp(-b_k y_t)
\propto
b_k^{n_k a_k}
\exp\left\{
-b_k\sum_{t:s_t=k}y_t
\right\}.
\]
Combining this term with the prior
\[
p(b_k)\propto b_k^{\zeta-1}\exp(-\xi b_k),
\]
we obtain
\[
p(b_k\mid a_k,y_1,\ldots,y_T,\boldsymbol{s})
\propto
b_k^{\zeta+n_k a_k-1}
\exp\left\{
-b_k\left(\xi+\sum_{t:s_t=k}y_t\right)
\right\}.
\]
After updating both parameters, the auxiliary mean parameter is simply recovered as
\[
\lambda_k=\frac{a_k}{b_k},
\]
to be used in the subsequent update of $a_k$.

\subsection{Circular part}

For the circular component we assume
\[
f_x(x_t\mid\mu_k,\kappa_k)
=
\frac{1}{2\pi I_0(\kappa_k)}
\exp\left\{
\kappa_k\cos(x_t-\mu_k)
\right\},
\]
where $\mu_k$ and $\kappa_k$ denote the mean direction and concentration.

\subsubsection*{Sampling of $\mu_k$}
We assign the von Mises prior
\[
\mu_k \sim \mathrm{VM}(\mu_0,\kappa_0).
\]
Then, the posterior distribution of $\mu_k$ is
\[
\mu_k\mid
\kappa_k,
\{x_t:s_t=k\}
\sim
\mathrm{VM}(\mu_{n_k},\kappa_{n_k}),
\]
where
\[
C_k
=
\kappa_k
\sum_{t:s_t=k}\cos(x_t)
+
\kappa_0\cos(\mu_0),
\]
\[
S_k
=
\kappa_k
\sum_{t:s_t=k}\sin(x_t)
+
\kappa_0\sin(\mu_0),
\]

\[
\mu_{n_k}
=
\operatorname{atan2}(S_k,C_k),
\qquad
\kappa_{n_k}
=
\sqrt{C_k^2+S_k^2}.
\]
In fact, conditionally on $\kappa_k$, the likelihood contribution involving $\mu_k$ is
\[
\prod_{t:s_t=k}
\exp\{\kappa_k\cos(x_t-\mu_k)\}
\propto
\exp\left\{
\kappa_k\sum_{t:s_t=k}\cos(x_t-\mu_k)
\right\}.
\]
Using
\[
\cos(x_t-\mu_k)=\cos x_t\cos\mu_k+\sin x_t\sin\mu_k,
\]
and combining with the prior
\[
p(\mu_k\mid\kappa_k)
\propto
\exp\{\kappa_0\cos(\mu_k-\mu_0)\},
\]
we obtain
\[
p(\mu_k\mid\kappa_k,\{x_t:s_t=k\})
\propto
\exp\{C_k\cos\mu_k+S_k\sin\mu_k\}.
\]
Since
\[
C_k\cos\mu_k+S_k\sin\mu_k
=
\kappa_{n_k}\cos(\mu_k-\mu_{n_k}),
\]
with
\[
\mu_{n_k}=\operatorname{atan2}(S_k,C_k),
\qquad
\kappa_{n_k}=\sqrt{C_k^2+S_k^2},
\]
the posterior is
\[
\mu_k\mid \kappa_k,\{x_t:s_t=k\}
\sim
\mathrm{VM}(\mu_{n_k},\kappa_{n_k}).
\]

\subsubsection*{Sampling of $\kappa_k$}
Conditionally on $\mu_k$, the likelihood contribution involving $\kappa_k$ is
\[
\prod_{t:s_t=k}
\frac{1}{I_0(\kappa_k)}
\exp\{\kappa_k\cos(x_t-\mu_k)\}
\propto
I_0(\kappa_k)^{-n_k}
\exp\left\{
\kappa_k
\sum_{t:s_t=k}
\cos(x_t-\mu_k)
\right\}.
\]
Combining this likelihood with the prior
\[
\pi(\kappa_k)
\propto
I_0(\kappa_k)^{-\rho}
\exp(-\chi\kappa_k),
\]
yields
\[
p(\kappa_k
\mid
\mu_k,
\{x_t:s_t=k\})
\propto
I_0(\kappa_k)^{-(\rho+n_k)}
\exp\left\{
-\kappa_k
\left(
\chi-
\sum_{t:s_t=k}
\cos(x_t-\mu_k)
\right)
\right\}.
\]
Defining
\[
m
=
\rho+n_k,
\]
and
\[
\beta_0
=
\frac{
\chi
-
\sum_{t:s_t=k}
\cos(x_t-\mu_k)
}
{\rho+n_k},
\]
we obtain
\[
p(\kappa_k
\mid
\mu_k,
\{x_t:s_t=k\})
\propto
I_0(\kappa_k)^{-m}
\exp(-m\beta_0\kappa_k).
\]
The conditional posterior of $\kappa_k$ therefore belongs to the Bessel--exponential family. Since its normalizing constant is not available in closed form, we sample $\kappa_k$ using the efficient shifted-Gamma rejection sampler of \citet{forbes2015fast}.

For large values of $\kappa_k$, the approximation
\[
I_0(\kappa_k)
\approx
\frac{e^{\kappa_k}}{\sqrt{2\pi\kappa_k}}
\]
shows that the target density is approximately Gamma. To obtain a proposal that remains accurate also for small values of $\kappa_k$, \citet{forbes2015fast} introduce the shifted-Gamma proposal
\[
x
\sim
\mathrm{Gamma}(mq+1,m\omega),
\qquad
\kappa_k=x-\varepsilon,
\qquad
\varepsilon>0,
\]
whose acceptance probability is based on
\[
g(\kappa_k;q,\omega,\varepsilon)
=
(\omega-\beta_0)\kappa_k
-
q\log(\kappa_k+\varepsilon)
-
\log I_0(\kappa_k).
\]
The proposal parameters are chosen as follows. Let
\[
r(\kappa)=\frac{I_1(\kappa)}{I_0(\kappa)}.
\]
First compute the working point
\[
\kappa_L
=
\frac{2}
{m\beta_0+\sqrt{2m+m^2\beta_0^2}},
\qquad
\kappa_U
=
\frac{2+1/m}
{(m+1)\beta_0+\sqrt{2m+1+m^2\beta_0^2}},
\]
\[
c_1
=
\frac12
+
\frac{1-1/(2m)}{2m},
\qquad
\kappa_0
=
(1-c_1)\kappa_L+c_1\kappa_U,
\]
and evaluate
$
i_0=I_0(\kappa_0)
$ and $
r_0=r(\kappa_0).
$
Next, define $
c_2
=
\frac{1}{4m}
-
\frac{2}{3\sqrt{m}},
$
and
\begin{equation}
\label{eq:omega}
\omega
=
\begin{cases}
\beta_0+1,
&
\beta_0\le c_2,
\\[6pt]
\beta_0+r_0+
\dfrac{1-r_0}
{1+40m(\beta_0-c_2)^2},
&
\text{otherwise.}
\end{cases}
\end{equation}
Then compute
\[
c_3
=
\frac{\log(i_0)/\kappa_0-\omega+\beta_0}
{\omega-\beta_0-r_0},
\]
\begin{equation}
\label{eq:eps}
\varepsilon
=
\frac{\kappa_0W_0(c_3e^{c_3})}
{c_3-W_0(c_3e^{c_3})},
\end{equation}
where $W_0(\cdot)$ denotes the principal branch of the Lambert $W$ function, defined as the inverse of the mapping $w\mapsto we^w$, i.e., $W_0(x)e^{W_0(x)}=x$.
Finally, set
\begin{equation}
\label{eq:q}
q
=
(\omega-\beta_0-r_0)
(\kappa_0+\varepsilon).
\end{equation}
Algorithm~\ref{alg:kappa-sampler} summarizes the shifted-Gamma rejection sampler used to update $\kappa_k$.

\begin{algorithm}[H]
\caption{Shifted-Gamma rejection sampler for $\kappa_k$ \citep{forbes2015fast}}
\label{alg:kappa-sampler}
\begin{algorithmic}[1]
\Require Parameters $m>0$, $\beta_0>-1$
\State Compute the working point $\kappa_0$ and set
\[
i_0=I_0(\kappa_0),
\qquad
r_0=\frac{I_1(\kappa_0)}{I_0(\kappa_0)}.
\]
\State Compute the parameters $\omega$, $\varepsilon$ and $q$ as in \eqref{eq:omega}, \eqref{eq:eps} and \eqref{eq:q}
\Repeat
\State Draw
$
x\sim
\mathrm{Gamma}(mq+1,m\omega)
$
truncated to $(\varepsilon,\infty)$
\State Set $\kappa_k=x-\varepsilon$
\State Draw $u\sim\mathcal{U}(0,1)$
\State Accept $\kappa_k$ if
\[
\frac{\log u}{m}
<
(\omega-\beta_0)(\kappa_k-\kappa_0)
-
q\log\left(
\frac{\kappa_k+\varepsilon}
     {\kappa_0+\varepsilon}
\right)
-
\log\left(
\frac{I_0(\kappa_k)}
     {i_0}
\right)
\]
\Until{$\kappa_k$ is accepted}
\State \Return $\kappa_k$
\end{algorithmic}
\end{algorithm}

\section{Additional simulation results}
\label{sec:sim}

In this section, we provide additional results from the simulation study.

Table~\ref{tab:Kselect} reports the distribution of the posterior number of states when the true number of states is \(K=4\). The most challenging scenario corresponds to the shortest time series \(T=100\). This result is expected, as reliable estimation of all state-specific parameters requires each latent state to be visited sufficiently often, which is less likely in short series. The accuracy of model selection improves substantially as the state persistence increases. Results for \(K=2\) are omitted since the correct number of states is recovered with probability exceeding 0.90 in all simulation settings.
\begin{table}[t]
\centering
\caption{Percentage of replications selecting each estimated number of states when the true number of states is $K=4$. Bold values indicate the highest selection probability for each simulation setting.}
\label{tab:Kselect}
\begin{tabular}{ccccccc}
\midrule
$\pi_{ii}$&$T$  & $\hat K=1$ & $\hat K=2$ & $\hat K=3$ & $\hat K=4$ & $\hat K=5$\\
\midrule
&100   & 10.5 & \textbf{47.0} & 42.5 & 0.0  & 0.0 \\
0.70&500   & 0.0  & 0.0  & 44.0 & \textbf{56.0} & 0.0 \\
&1\,000  & 0.0  & 0.0  & 8.0  & \textbf{90.5} & 1.5 \\[0.2em]

&100   & 2.5  & 41.0 & \textbf{52.0} & 4.5  & 0.0 \\
0.95&500   & 0.0  & 0.0  & 7.5  & \textbf{89.0} & 3.5 \\
&1\,000  & 0.0  & 0.0  & 8.5  & \textbf{88.5} & 3.0 \\
\bottomrule
\end{tabular}
\end{table}

Table~\ref{tab:par} reports the true parameter values together with the posterior median estimates for each emission parameter under the different combinations of \(K\), \(T\), and \(\pi_{ii}\). Values in parentheses denote the posterior coefficients of variation: as expected, the largest values are observed for angular means close to \(0\), since \(0\) and \(2\pi\) represent the same direction and the posterior distribution may therefore place mass around both values. 
Overall, the emission parameters are accurately recovered across all simulation settings. Estimation improves as $\pi_{ii}$ increases, while increasing the series length mainly reduces posterior uncertainty. 

\begin{table}[t]
\centering
\scriptsize
\setlength{\tabcolsep}{3pt}
\caption{Posterior median of the emission parameters, with the posterior coefficients of variation in parentheses. The column ``Par.'' denotes the parameter of interest, while the column ``True'' reports the parameter values used for data generation.}
\label{tab:par}
\resizebox{\textwidth}{!}{
\begin{tabular}{cclcccccc}
\toprule
&
&
&
\multicolumn{3}{c}{$\pi_{ii}=0.70$} &
\multicolumn{3}{c}{$\pi_{ii}=0.95$} \\
\cmidrule(lr){4-6}\cmidrule(lr){7-9}
$K$ & Par. & True &
$T=100$ & $T=500$ & $T=1000$ &
$T=100$ & $T=500$ & $T=1000$\\
\midrule

\multirow{8}{*}{$2$}

& $\mu_1$ & $0=2\pi$
& 6.198 (0.411)
& 0.032 (1.421)
& 6.260 (0.775)
& 6.219 (0.433)
& 0.036 (2.865)
& 6.261 (0.720)\\

&
\cellcolor{gray!10}$\mu_2$ &
\cellcolor{gray!10}$\pi/2$ &
\cellcolor{gray!10}1.348 (0.209) &
\cellcolor{gray!10}1.678 (0.124) &
\cellcolor{gray!10}1.348 (0.131) &
\cellcolor{gray!10}1.648 (0.161) &
\cellcolor{gray!10}1.511 (0.103) &
\cellcolor{gray!10}1.491 (0.092)\\

& $\kappa_1$ & $10$
& 6.398 (0.200)
& 8.767 (0.094)
& 9.392 (0.065)
& 8.108 (0.170)
& 7.794 (0.087)
& 8.613 (0.062)\\

&
\cellcolor{gray!10}$\kappa_2$ &
\cellcolor{gray!10}$0.5$ &
\cellcolor{gray!10}0.799 (0.243) &
\cellcolor{gray!10}0.469 (0.132) &
\cellcolor{gray!10}0.395 (0.100) &
\cellcolor{gray!10}0.923 (0.257) &
\cellcolor{gray!10}0.602 (0.117) &
\cellcolor{gray!10}0.467 (0.091)\\

& $a_1$ & $4$
& 3.997 (0.244)
& 4.445 (0.107)
& 4.408 (0.082)
& 3.127 (0.170)
& 4.002 (0.089)
& 4.503 (0.064)\\

&
\cellcolor{gray!10}$a_2$ &
\cellcolor{gray!10}$4$ &
\cellcolor{gray!10}3.833 (0.199) &
\cellcolor{gray!10}3.256 (0.087) &
\cellcolor{gray!10}4.639 (0.064) &
\cellcolor{gray!10}4.028 (0.218) &
\cellcolor{gray!10}3.378 (0.087) &
\cellcolor{gray!10}4.068 (0.061)\\

& $b_1$ & $2$
& 1.843 (0.237)
& 2.127 (0.106)
& 2.076 (0.079)
& 1.481 (0.183)
& 2.020 (0.093)
& 2.185 (0.067)\\

&
\cellcolor{gray!10}$b_2$ &
\cellcolor{gray!10}$8$ &
\cellcolor{gray!10}7.951 (0.217) &
\cellcolor{gray!10}6.368 (0.097) &
\cellcolor{gray!10}8.813 (0.070) &
\cellcolor{gray!10}8.824 (0.230) &
\cellcolor{gray!10}6.438 (0.095) &
\cellcolor{gray!10}7.701 (0.065)\\

\midrule

\multirow{16}{*}{$4$}

& $\mu_1$ & $3\pi/2$
& --
& 4.416 (0.029)
& 4.738 (0.021)
& 4.711 (0.089)
& 4.752 (0.027)
& 4.705 (0.018)\\

&
\cellcolor{gray!10}$\mu_2$ &
\cellcolor{gray!10}$\pi$ &
\cellcolor{gray!10}-- &
\cellcolor{gray!10}3.131 (0.046) &
\cellcolor{gray!10}3.161 (0.025) &
\cellcolor{gray!10}2.864 (0.087) &
\cellcolor{gray!10}3.061 (0.044) &
\cellcolor{gray!10}3.090 (0.024)\\

& $\mu_3$ & $\pi/2$
& --
& 1.543 (0.249)
& 1.500 (0.160)
& 1.451 (0.270)
& 1.910 (0.232)
& 1.438 (0.100)\\

&
\cellcolor{gray!10}$\mu_4$ &
\cellcolor{gray!10}$0=2\pi$ &
\cellcolor{gray!10}-- &
\cellcolor{gray!10}0.056 (1.365) &
\cellcolor{gray!10}0.034 (2.403) &
\cellcolor{gray!10}6.086 (0.529) &
\cellcolor{gray!10}6.206 (0.733) &
\cellcolor{gray!10}0.050 (1.124)\\

& $\kappa_1$ & $1$
& --
& 1.051 (0.153)
& 1.076 (0.134)
& 0.666 (0.308)
& 0.871 (0.123)
& 1.174 (0.100)\\

&
\cellcolor{gray!10}$\kappa_2$ &
\cellcolor{gray!10}$2$ &
\cellcolor{gray!10}-- &
\cellcolor{gray!10}1.533 (0.219) &
\cellcolor{gray!10}1.521 (0.132) &
\cellcolor{gray!10}2.167 (0.365) &
\cellcolor{gray!10}1.856 (0.177) &
\cellcolor{gray!10}1.765 (0.103)\\

& $\kappa_3$ & $0.5$
& --
& 0.671 (0.373)
& 0.783 (0.257)
& 0.675 (0.295)
& 0.503 (0.228)
& 0.535 (0.112)\\

&
\cellcolor{gray!10}$\kappa_4$ &
\cellcolor{gray!10}$10$ &
\cellcolor{gray!10}-- &
\cellcolor{gray!10}6.357 (0.136) &
\cellcolor{gray!10}7.784 (0.093) &
\cellcolor{gray!10}3.305 (0.312) &
\cellcolor{gray!10}6.306 (0.173) &
\cellcolor{gray!10}7.533 (0.105)\\

& $a_1$ & $4$
& --
& 3.076 (0.118)
& 3.586 (0.097)
& 2.964 (0.222)
& 3.714 (0.101)
& 4.741 (0.089)\\

&
\cellcolor{gray!10}$a_2$ &
\cellcolor{gray!10}$4$ &
\cellcolor{gray!10}-- &
\cellcolor{gray!10}3.447 (0.229) &
\cellcolor{gray!10}3.727 (0.128) &
\cellcolor{gray!10}2.350 (0.383) &
\cellcolor{gray!10}3.919 (0.165) &
\cellcolor{gray!10}3.613 (0.110)\\
& $a_3$ & $4$
& --
& 4.239 (0.211)
& 5.198 (0.203)
& 3.988 (0.219)
& 3.477 (0.143)
& 4.218 (0.077)\\

&
\cellcolor{gray!10}$a_4$ &
\cellcolor{gray!10}$4$ &
\cellcolor{gray!10}-- &
\cellcolor{gray!10}3.160 (0.139) &
\cellcolor{gray!10}3.537 (0.097) &
\cellcolor{gray!10}3.179 (0.339) &
\cellcolor{gray!10}3.748 (0.186) &
\cellcolor{gray!10}4.397 (0.108)\\

& $b_1$ & $8$
& --
& 5.727 (0.145)
& 7.198 (0.117)
& 5.709 (0.240)
& 7.385 (0.109)
& 9.048 (0.096)\\

&
\cellcolor{gray!10}$b_2$ &
\cellcolor{gray!10}$2$ &
\cellcolor{gray!10}-- &
\cellcolor{gray!10}1.649 (0.194) &
\cellcolor{gray!10}1.909 (0.112) &
\cellcolor{gray!10}1.161 (0.415) &
\cellcolor{gray!10}2.071 (0.168) &
\cellcolor{gray!10}1.723 (0.116)\\

& $b_3$ & $4$
& --
& 3.650 (0.223)
& 5.404 (0.238)
& 3.992 (0.233)
& 3.206 (0.153)
& 4.103 (0.081)\\

&
\cellcolor{gray!10}$b_4$ &
\cellcolor{gray!10}$1.33$ &
\cellcolor{gray!10}-- &
\cellcolor{gray!10}1.140 (0.143) &
\cellcolor{gray!10}1.210 (0.097) &
\cellcolor{gray!10}0.910 (0.362) &
\cellcolor{gray!10}1.205 (0.194) &
\cellcolor{gray!10}1.392 (0.112)\\

\bottomrule
\end{tabular}
}
\end{table}

\begin{figure}
    \centering
    \includegraphics[width=\linewidth]{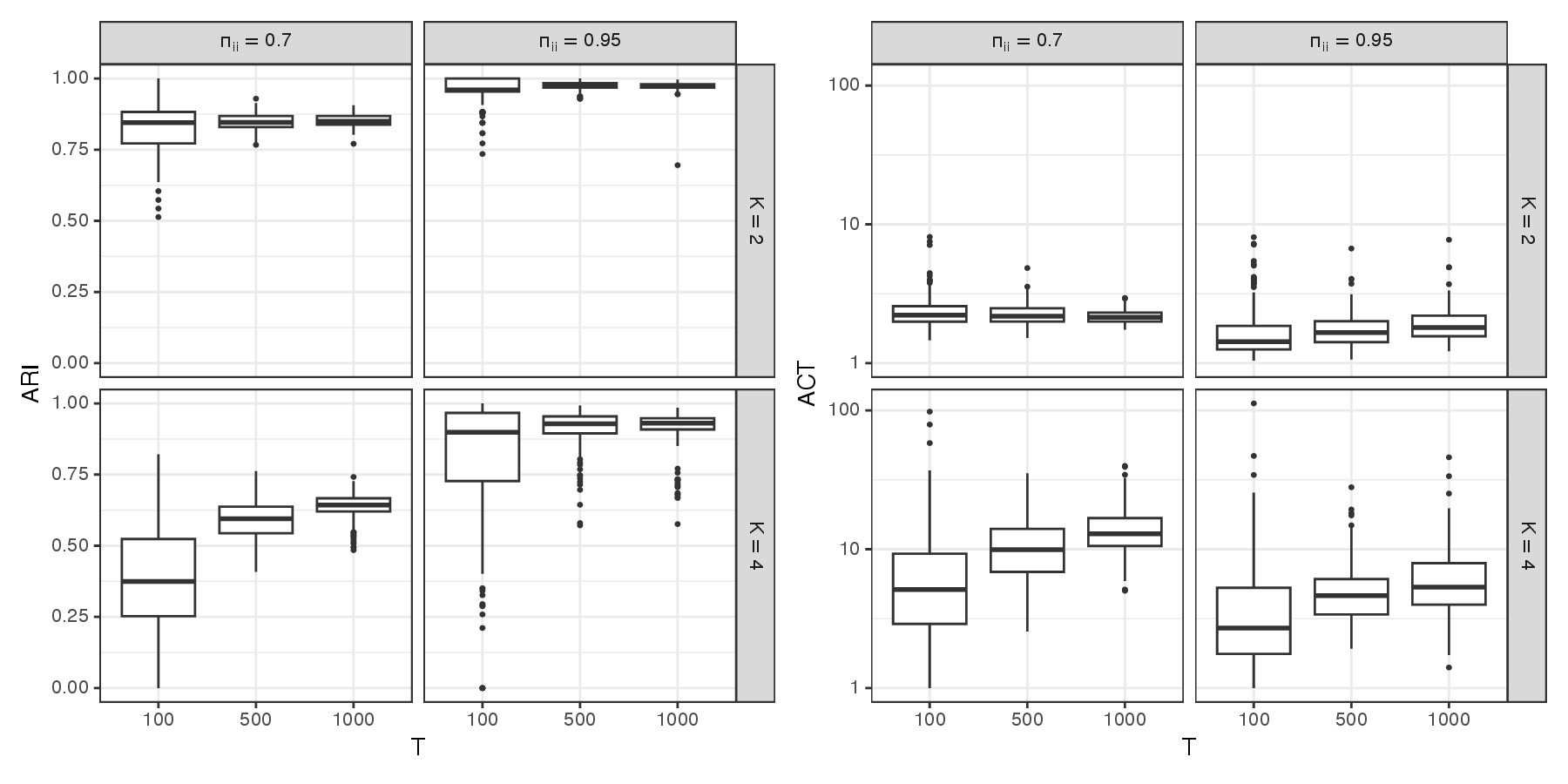}
    \caption{Boxplots of the adjusted Rand index (ARI, left) and the median autocorrelation time (ACT, right) obtained across simulation replicates. Results are reported for different time-series lengths $T$, with panels corresponding to different numbers of latent states $K$, and self-transition probabilities $\pi_{ii}$.}
    \label{fig:bac}
\end{figure}

Figure~\ref{fig:bac} displays boxplots of the ARI and ACT across the different combinations of \(K\), \(\pi_{ii}\), and \(T\). 
The ARI values indicate satisfactory recovery of the latent state sequence across all simulation settings, except in the most challenging scenario \(K=4, T=100\), as discussed above.
The ACT generally increases when \(K=4\).

\section{Additional results from the empirical analysis}
\label{sec:emp}

In this section, we provide additional figures and tables complementing the empirical analyses presented in Section~4 of the main paper.

\subsection{Drosophila melanogaster movement}

All of the 10 independent chains show $K=2$ as their posterior mode for the wild-type dataset, whereas 8 out of 10 converge to $K=3$ for the mutant dataset.

\begin{figure}[t]
    \centering

    \begin{subfigure}[t]{0.8\linewidth}
        \centering
        \caption{Wild-type}\includegraphics[width=\linewidth]{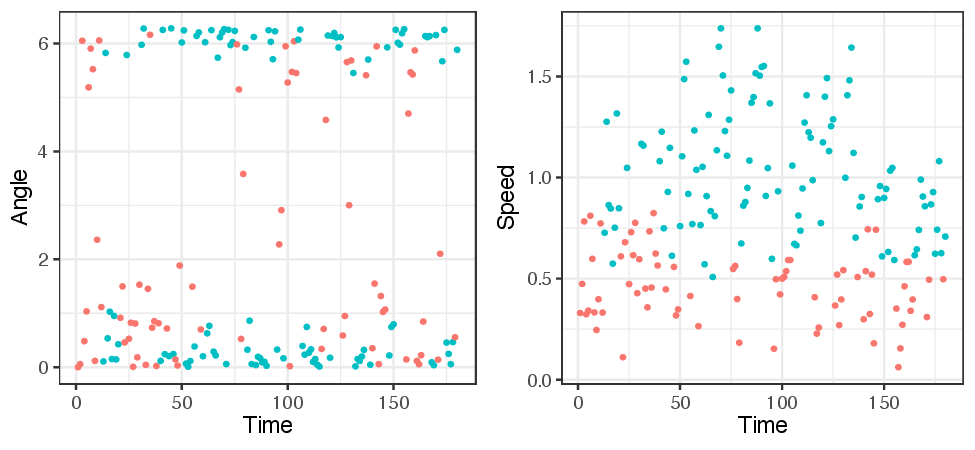}
        \label{fig:MAP_angle_speed_w}
    \end{subfigure}

    \vspace{0.2em}

    \begin{subfigure}[t]{0.8\linewidth}
        \centering
        \caption{Mutant}\includegraphics[width=\linewidth]{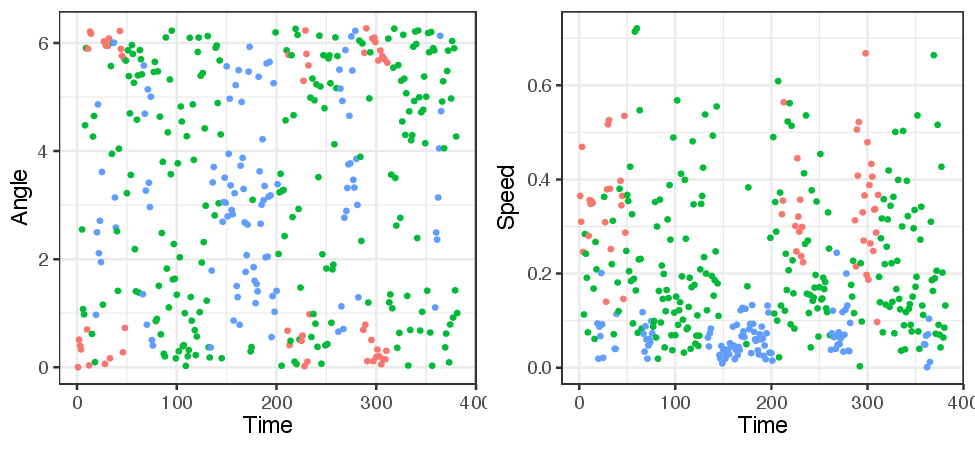}
        \label{fig:MAP_angle_speed_m}
    \end{subfigure}

    \caption{Observed circular direction and linear speed time series. Points are colored according to the maximum a posteriori inferred latent state.}
    \label{fig:MAP_angle_speed}
\end{figure}

Figure~\ref{fig:MAP_angle_speed} displays the observed movement directions and speeds for the wild-type and mutant \textit{Drosophila} datasets, with each observation colored according to its maximum a posteriori (MAP) inferred state. The wild-type larvae, Figure \ref{fig:MAP_angle_speed_w}, are characterized by two distinct behavioral states: one with fast locomotion, and turning angle highly concentrated around 0 (light blue), the other with slow locomotion and greater directional variability (red). 
In contrast, the mutant, Figure \ref{fig:MAP_angle_speed_m} exhibits three regimes, including an additional intermediate-speed state (green).

Table~\ref{tab:droso} reports posterior summaries of the state-specific emission parameters for both genotypes. For the wild type, the second state is characterized by both a substantially higher mean speed and a much stronger directional concentration than the first state. In contrast, the mutant displays three distinct regimes corresponding to fast, intermediate, and slow locomotion, with directional concentration decreasing as movement speed decreases. 
\begin{table}[ht]
\centering
\caption{Posterior summaries of the state-specific emission parameters. Entries report the posterior median, with the 95\% credible interval in parentheses.}
\label{tab:droso}

\begin{tabular}{lccc}
\toprule
\multicolumn{4}{c}{\textbf{Wild-type}}\\
\midrule
State & $\mu_k^{(w)}$ & $\kappa_k^{(w)}$ & $\lambda_k^{(w)}$\\
\midrule
1 &
\begin{tabular}{@{}c@{}}
0.294\\
{\scriptsize (0.058,\,0.755)}
\end{tabular}
&
\begin{tabular}{@{}c@{}}
1.521\\
{\scriptsize (0.895,\,2.133)}
\end{tabular}
&
\begin{tabular}{@{}c@{}}
0.505\\
{\scriptsize (0.414,\,0.595)}
\end{tabular}
\\[1ex]

2 &
\begin{tabular}{@{}c@{}}
0.067\\
{\scriptsize (0.004,\,6.279)}
\end{tabular}
&
\begin{tabular}{@{}c@{}}
7.095\\
{\scriptsize (5.058,\,10.484)}
\end{tabular}
&
\begin{tabular}{@{}c@{}}
1.024\\
{\scriptsize (0.918,\,1.151)}
\end{tabular}
\\
\midrule
\end{tabular}


\begin{tabular}{lccc}
\multicolumn{4}{c}{\textbf{Mutant}}\\
\midrule
State & $\mu_k^{(m)}$ & $\kappa_k^{(m)}$ & $\lambda_k^{(m)}$\\
\midrule
1 &
\begin{tabular}{@{}c@{}}
6.179\\
{\scriptsize (0.006,\,6.277)}
\end{tabular}
&
\begin{tabular}{@{}c@{}}
4.157\\
{\scriptsize (2.560,\,6.278)}
\end{tabular}
&
\begin{tabular}{@{}c@{}}
0.358\\
{\scriptsize (0.314,\,0.407)}
\end{tabular}
\\[1ex]

2 &
\begin{tabular}{@{}c@{}}
6.058\\
{\scriptsize (0.018,\,6.267)}
\end{tabular}
&
\begin{tabular}{@{}c@{}}
0.664\\
{\scriptsize (0.430,\,0.899)}
\end{tabular}
&
\begin{tabular}{@{}c@{}}
0.221\\
{\scriptsize (0.196,\,0.250)}
\end{tabular}
\\[1ex]

3 &
\begin{tabular}{@{}c@{}}
2.977\\
{\scriptsize (0.907,\,5.270)}
\end{tabular}
&
\begin{tabular}{@{}c@{}}
0.341\\
{\scriptsize (0.024,\,0.679)}
\end{tabular}
&
\begin{tabular}{@{}c@{}}
0.080\\
{\scriptsize (0.068,\,0.095)}
\end{tabular}
\\
\bottomrule
\end{tabular}

\end{table}

\begin{figure}[t]
    \centering
    \includegraphics[width=.8\linewidth]{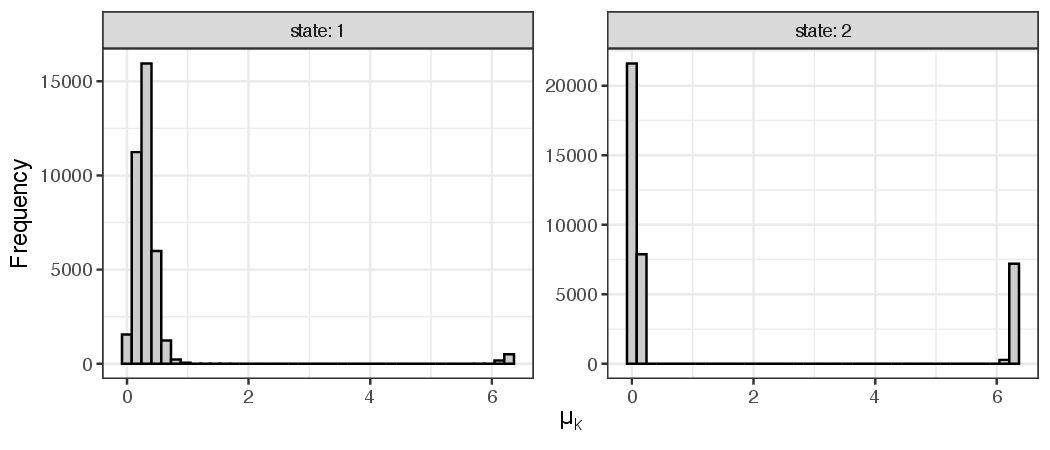}
    \includegraphics[width=.8\linewidth]{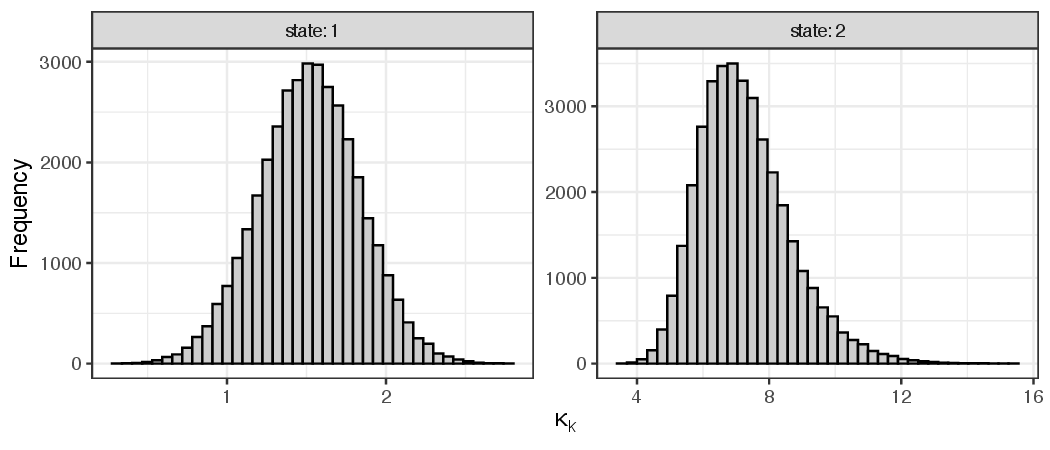}
    \includegraphics[width=.8\linewidth]{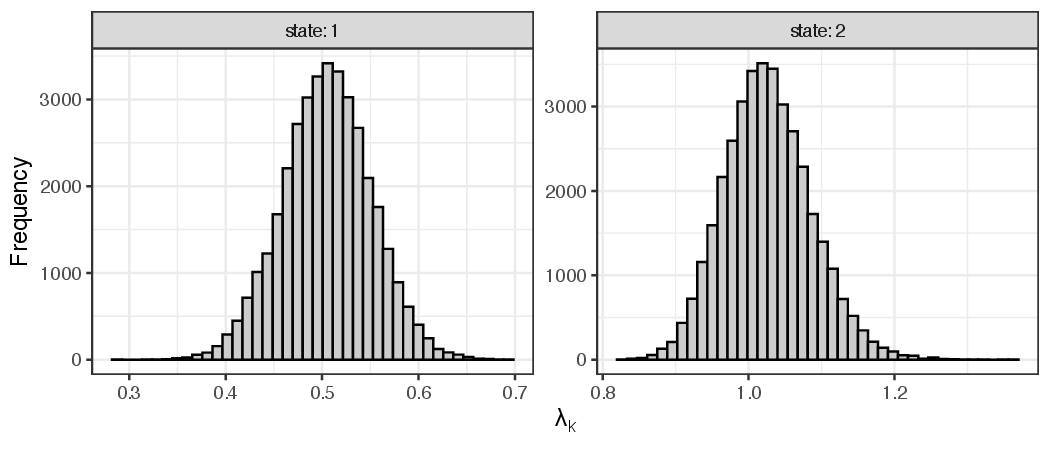}
    \caption{
    Posterior distributions of the state-specific emission parameters $\boldsymbol{\mu}^{(w)}$, $\boldsymbol{\kappa}^{(w)}$, $\boldsymbol{\lambda}^{(w)}$.}
    \label{fig:par_w}
\end{figure}

Figure ~\ref{fig:par_w} shows the posterior distributions 
of the state-specific parameters for the wild-type dataset, which clearly distinguish between the two states. 

\begin{figure}[t]
    \centering
    \includegraphics[width=.8\linewidth]{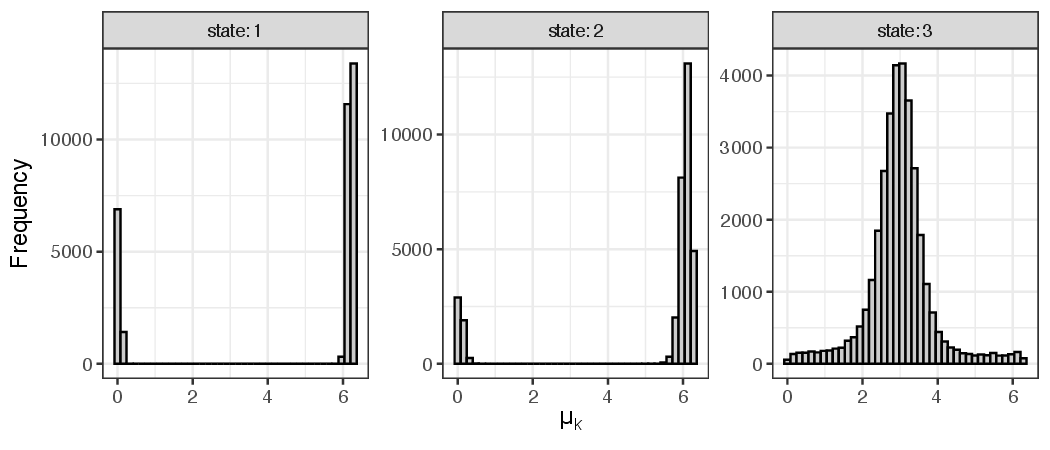}
    \includegraphics[width=.8\linewidth]{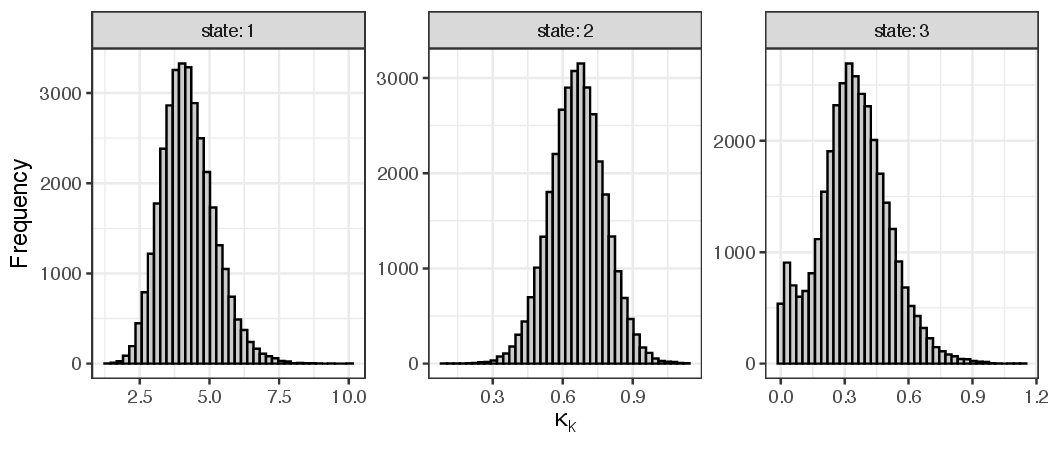}
    \includegraphics[width=.8\linewidth]{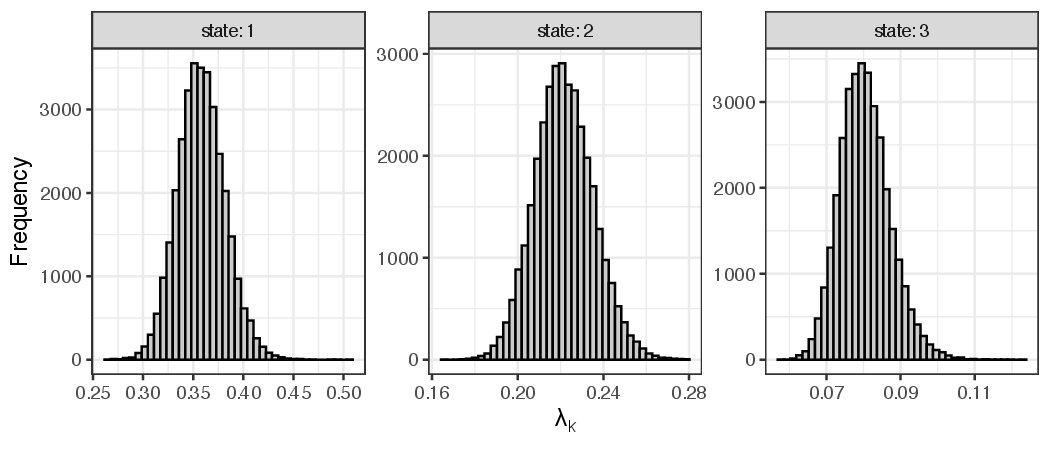}
    \caption{
    Posterior distributions of the state-specific emission parameters $\boldsymbol{\mu}^{(m)}$, $\boldsymbol{\kappa}^{(m)}$, $\boldsymbol{\lambda}^{(m)}$.}
    \label{fig:par_m}
\end{figure}

Figure \ref{fig:par_m} shows the posterior distributions of the emission parameters for the mutant dataset, clearly distinguishing three behavioral states.

\subsection{Wind dynamics}

Out of the 10 independent chains, 6 select $K=5$ as the posterior mode.

\begin{figure}
    \centering
    \includegraphics[width=.8\linewidth]{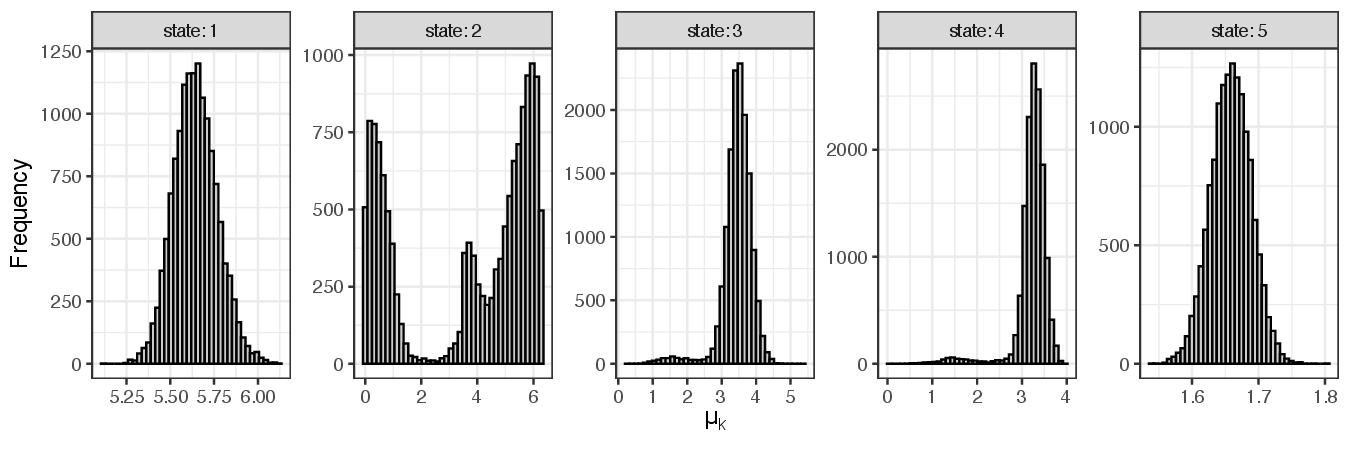}
    \includegraphics[width=.8\linewidth]{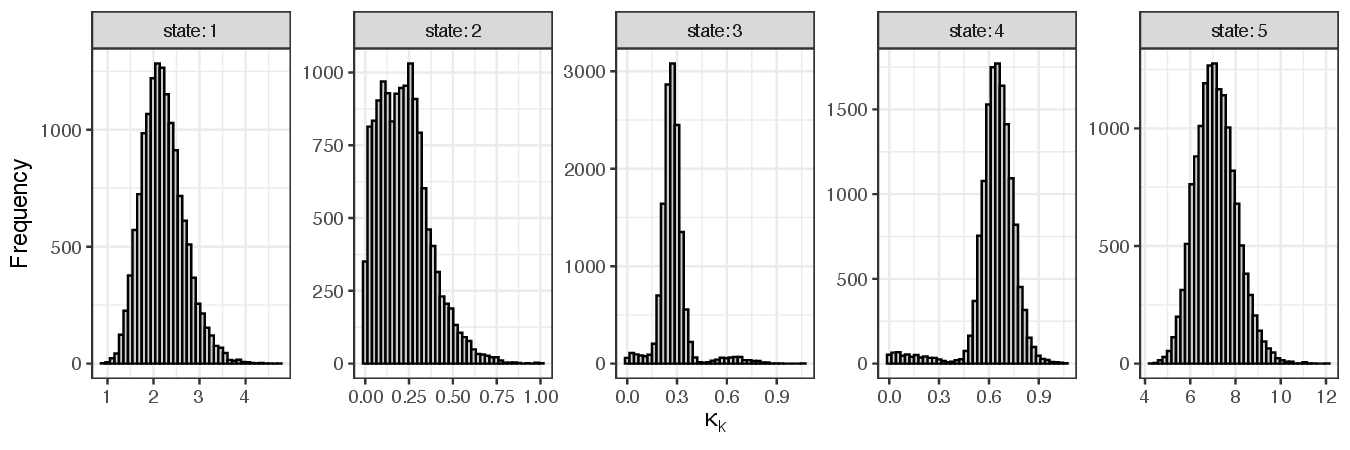}

    \includegraphics[width=.8\linewidth]{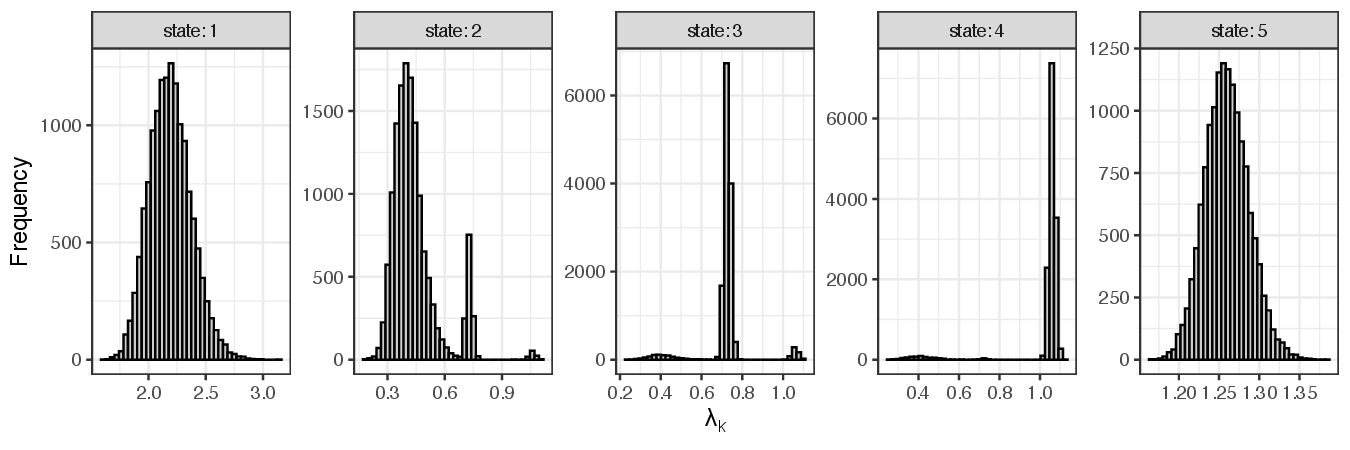}
    
    \caption{Posterior distribution of the state-specific mean direction  $\mu_k$, concentration $\kappa_k$, and average speed $\lambda_k$, $k=1,\ldots,5$.}
    \label{fig:mu_wind}
\end{figure}

Figure~\ref{fig:mu_wind} displays the posterior distributions for $\mu_1,\ldots,\mu_5$, and results are consistent with the state characterization discussed in the main paper. 

\clearpage
\bibliographystyle{elsarticle-harv}
\bibliography{biblio_short}

@article{Harvey2023,
	author = {Harvey, Andrew and Palumbo, Dario},
	journal = {Journal of Time Series Analysis},
	journal3 = {J. Time Ser. Anal.},
	number = {4},
	pages = {374--392},
	publisher = {John Wiley \& Sons, Ltd},
	title = {Regime switching models for circular and linear time series},
	volume = {44},
	year = {2023},
	year1 = {2023}
    }

@article{beal2001infinite,
	author = {Beal, Matthew and Ghahramani, Zoubin and Rasmussen, Carl},
	journal = {Adv. Neural Inf. Process. Syst.},
	pages = {577--584},
	title = {The infinite hidden {M}arkov model},
	volume = {14},
	year = {2001}}

@article{chib1996calculating,
	author = {Chib, Siddhartha},
	journal = {J. Econom.},
	number = {1},
	pages = {79--97},
	publisher = {Elsevier},
	title = {Calculating posterior distributions and modal estimates in {M}arkov mixture models},
	volume = {75},
	year = {1996}}

@article{cortese2026comparison,
	author = {Cortese, Federico P and Rossini, Luca},
	journal = {Comput. Stat. Data Anal.},
	pages = {108440},
	publisher = {Elsevier},
	title = {A comparison between initialization strategies for the infinite hidden {M}arkov model},
	year = {2026}}

@article{ferguson1973bayesian,
	author = {Ferguson, Thomas S},
	journal = {Ann. Stat.},
	number = {2},
	pages = {209--230},
	publisher = {JSTOR},
	title = {A {B}ayesian analysis of some nonparametric problems},
	volume = {1},
	year = {1973}}

@article{forbes2015fast,
	author = {Forbes, Peter GM and Mardia, Kanti V},
	journal = {J. Stat. Comput. Simul.},
	number = {13},
	pages = {2693--2701},
	publisher = {Taylor \& Francis},
	title = {A fast algorithm for sampling from the posterior of a von {M}ises distribution},
	volume = {85},
	year = {2015}}

@incollection{Geweke1992,
	address = {Oxford, UK},
	author = {Geweke, John},
	booktitle = {Bayesian Statistics 4},
	editor = {Bernardo, J. M. and Berger, J. O. and Dawid, A. P. and Smith, A. F. M.},
	pages = {169--193},
	publisher = {Clarendon Press},
	title = {Evaluating the Accuracy of Sampling-Based Approaches to Calculating Posterior Moments},
	year = {1992}}

@article{gower1971general,
	author = {Gower, John C},
	journal = {Biometrics},
	pages = {857--871},
	publisher = {JSTOR},
	title = {A general coefficient of similarity and some of its properties},
	year = {1971}}

@article{holzmann2006hidden,
	author = {Holzmann, Hajo and Munk, Axel and Suster, Max and Zucchini, Walter},
	journal = {Environ. Ecol. Stat.},
	number = {3},
	pages = {325--347},
	publisher = {Springer},
	title = {Hidden Markov models for circular and linear-circular time series},
	volume = {13},
	year = {2006}}

@article{lagona2015hidden,
	author = {Lagona, Francesco and Picone, Marco and Maruotti, Antonello and Cosoli, Simone},
	journal = {Stoch. Environ. Res. Risk Assess.},
	number = {2},
	pages = {397--409},
	publisher = {Springer},
	title = {A hidden Markov approach to the analysis of space--time environmental data with linear and circular components},
	volume = {29},
	year = {2015}}

@article{miller2019fast,
	author = {Miller, Jeffrey W},
	journal = {J. Comput. Graph. Stat.},
	number = {2},
	pages = {476--480},
	publisher = {Taylor \& Francis},
	title = {Fast and accurate approximation of the full conditional for gamma shape parameters},
	volume = {28},
	year = {2019}}

@article{sethuraman1994constructive,
	author = {Sethuraman, Jayaram},
	journal = {Stat. Sin.},
	number = {2},
	pages = {639--650},
	publisher = {JSTOR},
	title = {A constructive definition of {D}irichlet priors},
	volume = {4},
	year = {1994}}

@article{suster2003targeted,
	author = {Suster, Maximiliano L and Martin, Jean-Rene and Sung, Carl and Robinow, Steven},
	journal = {J. Neurobiol.},
	number = {2},
	pages = {233--246},
	publisher = {Wiley Online Library},
	title = {Targeted expression of tetanus toxin reveals sets of neurons involved in larval locomotion in {D}rosophila},
	volume = {55},
	year = {2003}}

@article{teh2006hierarchical,
	author = {Teh, Yee Whye and Jordan, Michael I and Beal, Matthew J and Blei, David M},
	journal = {J. Am. Stat. Assoc.},
	number = {476},
	pages = {1566--1581},
	publisher = {Taylor \& Francis},
	title = {Hierarchical {D}irichlet processes},
	volume = {101},
	year = {2006}}

@inproceedings{van2008beam,
	author = {Van Gael, Jurgen and Saatci, Yunus and Teh, Yee Whye and Ghahramani, Zoubin},
	booktitle = {Proc. 25th Int. Conf. Mach. Learn.},
	pages = {1088--1095},
	title = {Beam sampling for the infinite hidden {M}arkov model},
	year = {2008}}

@book{zucchini:2017,
	address = {Boca Raton, FL},
	author = {Zucchini, Walter and MacDonald, Iain L. and Langrock, Roland},
	publisher = {CRC Press},
	title = {Hidden {M}arkov Models for Time Series: An Introduction Using {R}},
	year = {2017}}

@article{lagona2019copula,
	author = {Lagona, Francesco},
	journal = {Stat. \& Prob. Letters},
	pages = {16--22},
	publisher = {Elsevier},
	title = {Copula-based segmentation of cylindrical time series},
	volume = {144},
	year = {2019}}

\end{document}